\documentclass{appolb}
\pdfoutput=1
\usepackage{epsfig}
\usepackage{graphicx}
\begin{document}
\title{Dark Matter Experimental Overview %
\thanks{Presented at the Cracow Epiphany Conference on Physics in Underground
Laboratories and Its Connection with LHC, Cracow, Poland, January 5 - 8, 2010.
}%
}
\author{Andrzej M. Szelc
\address{Laboratori Nazionali del Gran Sasso, 67010 Assergi L’Aquila, Italy}
 \address{                             and}
\address{The H. Niewodnicza\'nski Institute of Nuclear Physics PAN
              Radzikowskiego 152, Krakow, Poland}
}
\maketitle
\begin{abstract}
Dark Matter is one of the most intriguing riddles of modern astrophysics. The Standard Cosmological Model implies that only 4.5\% of the mass-energy of the Universe is baryonic matter and the remaining 95\% is unknown. Of this remainder, 22\% is expected to be Dark Matter - an entity that behaves like ordinary matter gravitationally but has not been yet observed in particle physics experiments and is not foreseen by the Standard Particle Model. It is expected that Dark Matter can be found in halos surrounding galaxies, the Milky Way among them, and it is hypothesized that it exists in the form of massive, weakly interacting particles i.e. WIMPs. A large experimental effort is being conducted to discover these elusive particles either directly, in underground laboratories, or indirectly, using experiments which search for decay or annihilation products of such particles in the night sky. This document aims to give a review of the status and recent results of selected Dark Matter searches.

\end{abstract}
  
\section{Introduction}

During the last two decades cosmology has seen an immense development. Some even say that it is only recently that it has become a science. However, there is still a certain paradox in the fact that we can measure cosmological parameters with an accuracy of few percent \cite{WMAP_recent} and yet we do not exactly know what constitutes 95\% of the mass-energy of the Universe. The Standard Cosmological Model \cite{SCM}, which describes our idea of how the Universe is constructed, is also called the Concordance Model since it is a result of many very diverse astrophysical and cosmological observations and measurements, such as the observations of the Cosmic Microwave Background \cite{WMAP}, the Supernovae type Ia \cite{SNIA} and the Large Scale Structure of the Universe \cite{SDSS} to name only a few. This model states that not more than $4.6$\% of the mass-energy of the Universe can be baryonic matter \cite{BBN_recent}. This limit is a result of the measurements of the abundances of light nuclei in the Universe \cite{L_nuclei}, which all converge on a value of the relic blackbody photon density parameter $\eta$ 
 which can be converted using the theory of nucleosynthesis into the total quantity of baryons in the Universe. The result is that ordinary matter cannot constitute more than 5\% of the so called critical density $\rho_c$. On the other hand, observations of the Cosmic Microwave Background, specifically the position of the first peak in the CMB power spectrum \cite{WMAP_recent}, lead to the conclusion that the Universe has a flat geometry which requires that the total mass-energy in the Universe be very close or equal to $\rho_c$. This means that in our mass-energy budget we are missing about 95\% of the $\rho_c$ or, in other words, roughly 20 times more than the matter we know and understand. Out of the missing mass-energy 73\% is assumed to be Dark Energy, a substance which leads to the acceleration of the expansion of the Universe, discovered when observing the SuperNovae type 1A and their emission spectra \cite{SNIA}. The remaining 22\% is ascribed to Dark Matter - an entity that interacts gravitationally like ordinary matter, but cannot be baryonic matter. Since Dark Matter probably does not consist of something that is known to us, its discovery will obviously lead to new physics and hopefully to a rapid development of our knowledge of the Universe and particle physics.

\subsection{Astronomical Evidence for Dark Matter}

Currently the success of the Concordance Model in describing our Universe is viewed as the strongest proof of the existence of Dark Matter. However, the idea that there is more in the Universe than we can account for has been present in astrophysics for the larger part of the former century. Already in the 1930s Fritz Zwicky noticed that the dynamics of the Coma Cluster suggests that there is more matter present in the system than one would expect from luminous matter \cite{Zwicky}. Another important observation suggesting a surplus of matter were the measurements of the galactic rotation curves \cite{Rubin} which exhibited a behavior incompatible with the hypothesis that the gravitational mass in galaxies is present in the galactic disk. On the contrary, the shape of the rotation curves suggested that the larger part of the mass in galaxies is distributed in a spherical halo. It is expected that the Milky Way should be no exception. Quite possibly, the most spectacular evidence for the existence of Dark Matter is the observation of colliding galaxy clusters named the Bullet Cluster \cite{Clowe}. The combination of optical and x-ray imaging as well as weak gravitational lensing led to the constatation that the main part of the cluster mass is weakly interacting contrary to the ionized hydrogen gas known to be the largest baryonic mass component in clusters, thus suggesting that the mass of galaxy clusters is a result of the presence of non-baryonic Dark Matter.

\subsection{Dark Matter Candidates}

Currently, the most popular candidate for Dark Matter is the so called WIMP (Weakly Interacting Massive Particles). Its seductivity lies in the fact that the expansions of the particle standard model predict a particle that could play the role of the WIMP. For example in Supersymmetry the Lightest Supersymmetric Particle (LSP) and for Kaluza-Klein extra dimension theories the Lightest Kaluza-Klein Particle (LKP). The standard theoretical WIMP candidate is the SuperSymmetric Neutralino particle, however there are many other possibilities both in SuperSymmetry and outside of it. Some of these other viable theoretical candidates are discussed with more detail in the corresponding theory review talk \cite{Mammbrini}. Another enticing property of the WIMP scenario is the idea that if we assume that the WIMP is a thermal relic of the Big Bang and assume it has very reasonable properties - a mass of $\sim$100GeV/c$^2$ and an interaction cross-section close to that of the weak interactions - the abundance of these relics would be very close to the current abundance of Dark Matter \cite{Kolb}. This is sometimes called the WIMP miracle.

Apart from the WIMP scenario, another DM candidate has good theoretical motivation, namely the solution of the strong CP problem. To theoretically explain the fact that CP violation is not observed in strong interactions, even though there is no principle that would forbid it, an extra scalar field is introduced which corresponds to a particle dubbed the axion. For a range of parameters the axions could solve the Dark Matter riddle, however since they are not thermal relics they are expected to have a mass that is much smaller than that expected in the WIMP scenario, in the range between  10$^{−6}$ to 3$\cdot$10$^{−2}$ eV/c$^2$ \cite{62}.

Recently the hypothesis that Dark Matter can interact inelastically \cite{Weiner} has been gaining some support in the community, however it will not be discussed here in detail due to lack of space.

\subsection{Plan}
In this paper we will try to outline the current experimental situation in Dark Matter detection both in the indirect and direct search experiments. The aim is to give a broad idea of the progress of the field, rather than an in-depth detailed, technical description of each experiment. Even so, due to the limited space it is not possible to represent all that is happening in this very interesting field and so some omissions had to be made. To begin, the current situation will be briefly signalled, followed by the most recent results of the indirect detection experiments. Next a presentation of the experimental techniques in direct detection together with sample experiments will be reported. Experiments searching for the diurnal modulation and axions will be mentioned very briefly, followed by the prospects for the nearest future. 

\section{Dark Matter Detection}

\begin{figure}[htp]
 \centering
 \includegraphics[width=60mm]{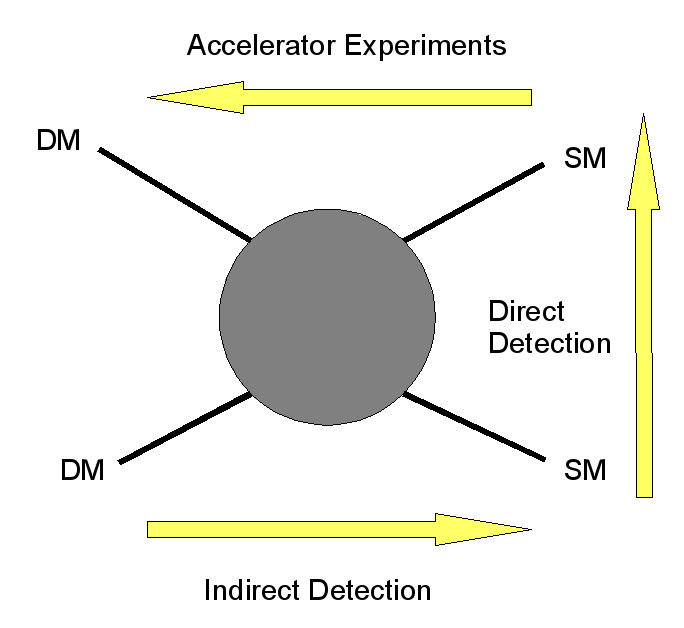}
 \caption{A diagram showing the principles of Dark Matter detection. ``DM'' signifies a Dark Matter particle and ``SM'' a known, Standard Model, one. Description in the text. }
 \label{fig:detection}
\end{figure}

Dark Matter detection is a difficult and challenging task. It can be approached in a variety of ways, which can be described by looking at Fig. \ref{fig:detection} from different angles. The standard, left to right mode, describes the idea behind indirect detection, where we assume that the Dark Matter particles annihilate and produce known standard model particles. Apart from this, indirect detection experiments may be able to register decays of heavy Dark Matter particles if they would be unstable. Fig. \ref{fig:detection} is of course a general idea and the devil lies in the details. First of all, the known particles must be extracted from an enormous background of cosmic ray particles arriving at the solar neighborhood from the galactic center and other directions, and determining this background is in itself a very challenging task. Even if some surplus particles are observed one must remember that the energy spectrum of the particles observed by us may have been somehow transformed during its voyage towards the Earth (for a more detailed discussion see \cite{Mammbrini} ), so it is again difficult to draw definite conclusions about the models that may have caused the observed energy surplus. On the other hand, it does give a rather large margin in which theoretical models can dwell. 

Looking at Fig. \ref{fig:detection} from right to left represents the idea of collider experiments, where we try to collide and annihilate known particles in order to obtain new, unknown, ones. It is quite probable that if new particles are indeed found in the accelerator experiments they will most probably be likely candidates for Dark Matter or at least lead toward a particle model in which Dark Matter candidates should be present. However, whether thess particles are really Dark Matter is a question that will have to be confirmed separately, probably by direct detection experiments. Another problem is that an actual WIMP candidate will be hard to find in the LHC, since its signature will most likely be that of missing energy. The topic of DM in accelerators is covered more thouroughly in \cite{DM_LHC}.
 
Lastly, Fig.\ref{fig:detection} can describe direct detection experiments if viewed from bottom to top. In this case the experiments look for an interaction where the WIMP scatters off of a detector nucleus transferring some of its kinetic energy. This recoil energy of the detector atom is what can be registered in the detector. However, there are many challenges connected to this task, since it is expected that these energies will be very small and the interactions very rare. So, to extract these interactions from background present in a detector will be a difficult task.

\subsection{Current Situation}
The last two years in Dark Matter searches have been what one can call interesting times. Experimental results have surfaced, which as of yet cannot provide definitive proof of the existence of Dark Matter, yet give, in the opinion of the author, a tantalizing sensation that something may be just around the corner. What is more, these results are not easily reconciled theoretically, thus creating a large stir in the theoretical community with literally hundreds of papers on the subject being published on the arxiv.org preprint archive.

The big commotion began with the announcement of a positron surplus by the PAMELA satellite observatory in August 2008 \cite{PAMELA}.  This was shortly followed by the publication of the observation of an excess of e$^-$ by the ATIC balloon borne observatory in November 2008 \cite{ATIC}. In May 2009 the situation became a bit more complicated when the FERMI-LAT observatory did see an overabundance of e$^+ +$e$^-$ \cite{FERMI}, however this excess was not in agreement with the ATIC observation. Apart from that, FERMI-LAT did not confirm the $\gamma$ ray overabundance seen with the EGRET detector \cite{EGRET}, which was considered a possible indirect Dark Matter signal before the PAMELA result. 

Last but not least, on December 17th 2009 CDMS II announced their analysis of their total data sample in which they observed two WIMP candidate events, with an expected background of one \cite{CDMS}. This was however not enough to claim discovery. One should also not forget the DAMA and DAMA-LIBRA result which is present on the market for almost a decade now\cite{DAMA} and which is a claim of the observation of Dark Matter using the so-called annual modulation effect. 

All in all, it would seem that there is more and more experimental evidence hinting at a discovery, even if none of the observations is yet able to pinpoint the Dark Matter particle. However, the experimental hints that we do have do not agree very well in the scope of our theoretical expectations. So there is still a lot of effort to be done in both the experimental and theoretical fields. 

\section{Indirect Detection}
As mentioned before, the indirect detection experiments try to focus on detecting the residues of decays or annihilations of Dark Matter particles. These can include antimatter and ordinary particles coming from annihilation into particle-antiparticle pairs (including neutrinos) or $\gamma$-radiation coming from decays and annihilations. The signals should most likely by observed from regions where a larger density of Dark Matter particles is expected, which could increase the chance of an annihilation event - this could be the Earth and the Sun for neutrino observations and the center of the galaxy for other particles. In this work we will concentrate on the experiments looking at the galactic center, since this is where most of the recent results come from. 

\subsection{PAMELA}

\begin{figure}[htp]
 \centering
 
 \includegraphics[width=50mm]{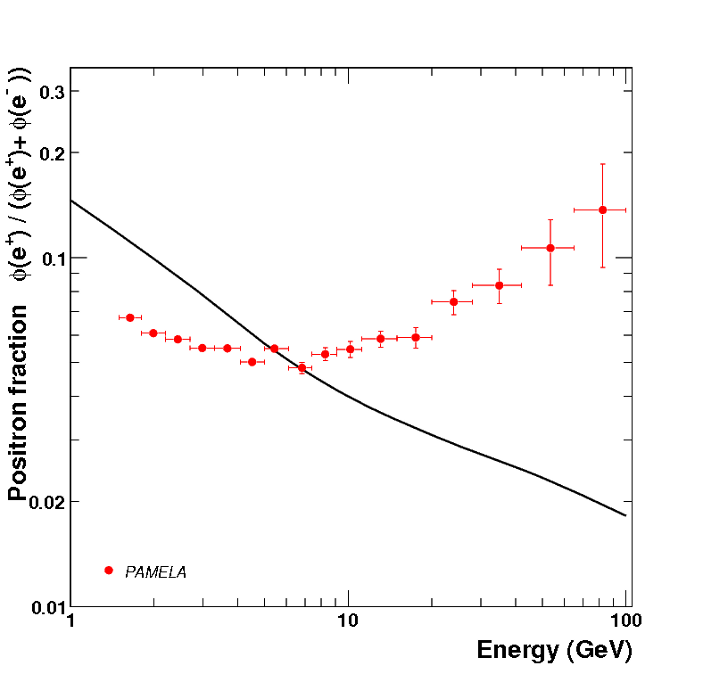}
 \includegraphics[width=50mm]{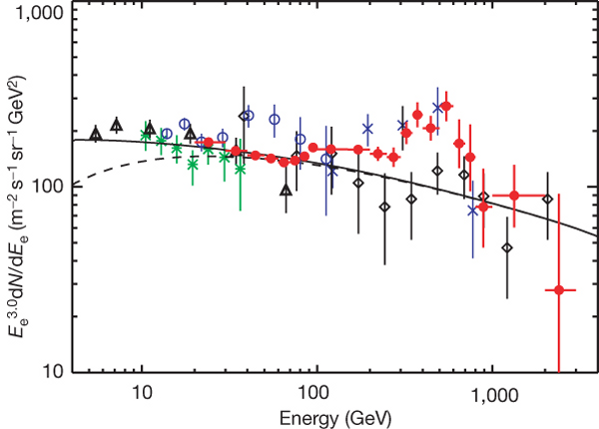}
\includegraphics[width=50mm,height=40mm]{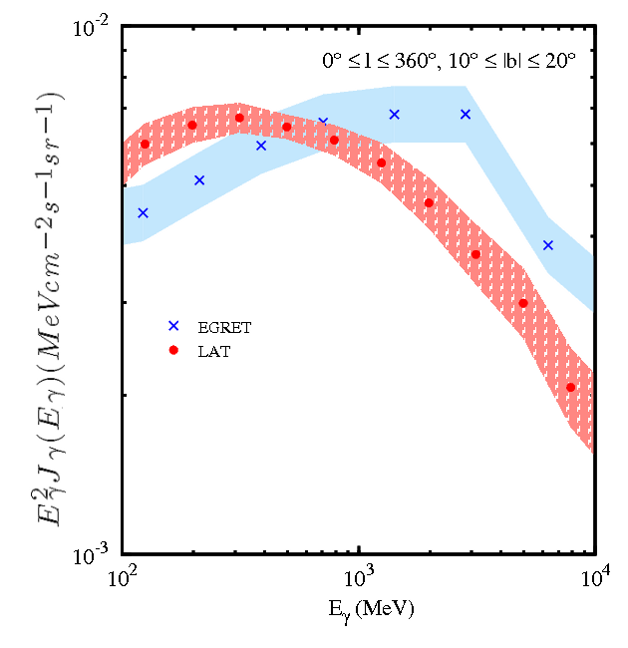}
 \includegraphics[width=50mm,,height=40mm]{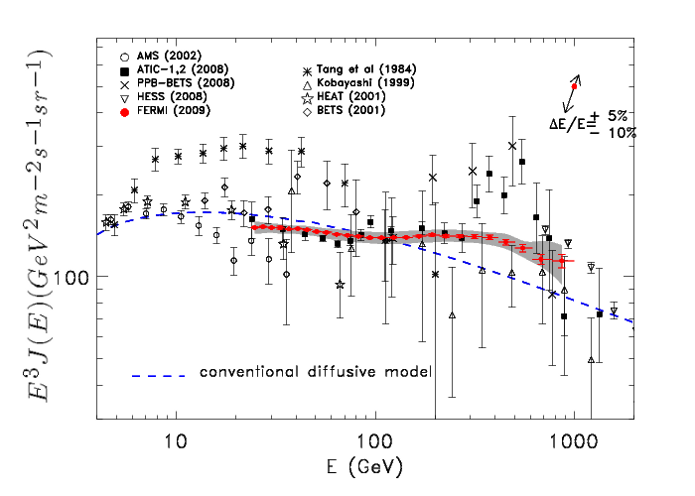}
 \caption{The  recent results of the indirect detection experiments. Upper left, the PAMELA e$^+$/(e$^+ +$e$^-$) overabundance \cite{PAMELA}, upper right is the ATIC spectrum with the e$^-$ surplus \cite{ATIC}, lower left the comparison of the EGRET $\gamma$ overabundance with the FERMI-LAT results \cite{FERMI-EGRET} and lower right the FERMI-LAT e$^+$ + e$^-$ spectrum \cite{FERMI}.}
 \label{fig:indirect}
\end{figure}

The PAMELA experiment is a satellite observatory equipped with
 a Time Of Flight detector, a calorimeter, a neutron detector
  and a spectrometer. It is set to observe charged particles in the cosmic rays as well as from the Sun. In their data run the detector observed an excess of e$^+$/(e$^+ +$e$^-$) \cite{PAMELA} with respect to the particles expected from recognized sources inside our galaxy \cite{GALPROP}, see Fig. \ref{fig:indirect} (upper left). This surplus is inconsistent with a standard WIMP scenario, since an overabundance in e$^+$ should be accompanied by an excess in antiprotons and this was not observed \cite{PAMELA_prots80}. This caused a large stir in the
  scientific community, exemplified by the fact that at the time of this talk, the PAMELA article has been cited more
  than 400 times. The discrepancy between positron and anti-protons gave rise to a family of leptophilic Dark Matter and other scenarios, but it is also compatible with the e$^+$/(e$^+ +$e$^-$) surplus being generated by standard galactic sources.


\subsection{ATIC}

The ATIC collaboration has operated balloon borne experiments in the Antarctic and has published their measurements of the e$^-$ spectrum not long after PAMELA.  Similarly, they observed a surplus of e$^-$ compared with expectations, which exhibited a peculiar structure in the region between  300 and 700 GeV \cite{ATIC}, see Fig. \ref{fig:indirect} (upper right). After the publication another run was included into the analysis and the preliminary results  seem to confirm the obtained result \cite{ATIC-unofficial}. Again, the ATIC spectrum would suggest a rather massive Dark Matter particle, rather hard to reconcile with the standard WIMP model.

\subsection{FERMI-LAT}

 The main scope of the FERMI Large Area Telescope (LAT) was to create the $\gamma$ map of our sky. It was launched June 2008.
One of the objectives, was to test the observations of the EGRET satellite, which observed a surplus of $\gamma$ rays compatible with the WIMP scenario \cite{EGRET}. FERMI-LAT was able to acquire a much larger statistic and did not confirm the EGRET excess, see Fig. \ref{fig:indirect} (bottom left). 

The FERMI-LAT detector was also used to look at charged particles, however since it was constructed to look at photons it is not able to discern electrons from positrons. Nevertheless, an overabundance of e$^+$ + e$^-$ was registered see Fig. \ref{fig:indirect} (bottom right). However, the FERMI-LAT result is not compatible with the ATIC observations.


\subsection{Nearest Future}

Even though the indirect detection experiments have recently provided very interesting results and several detectors
seem to see something above the predicted
  background, they are far from being conclusive. First of all their results
  are not fully consistent amongst   themselves i.e. FERMI and
  EGRET, FERMI and ATIC. They are also difficult to reconcile with
  the standard WIMP scenario Dark Matter. This has caused an enormous rise in Dark Matter model building.
There is still a chance, that all these observations could be explained by standard astrophysical
  phenomena, like pulsars or such. Definitely, more data is needed - fortunately
  FERMI and PAMELA are still taking data. In the meantime, the PLANCK Microwave Background observatory has also arrived in orbit, and the AMS detector
  should be launched in the summer of 2010.

\section{Direct Detection}

\begin{figure}[htp]
 \centering
 
\includegraphics[width=50mm]{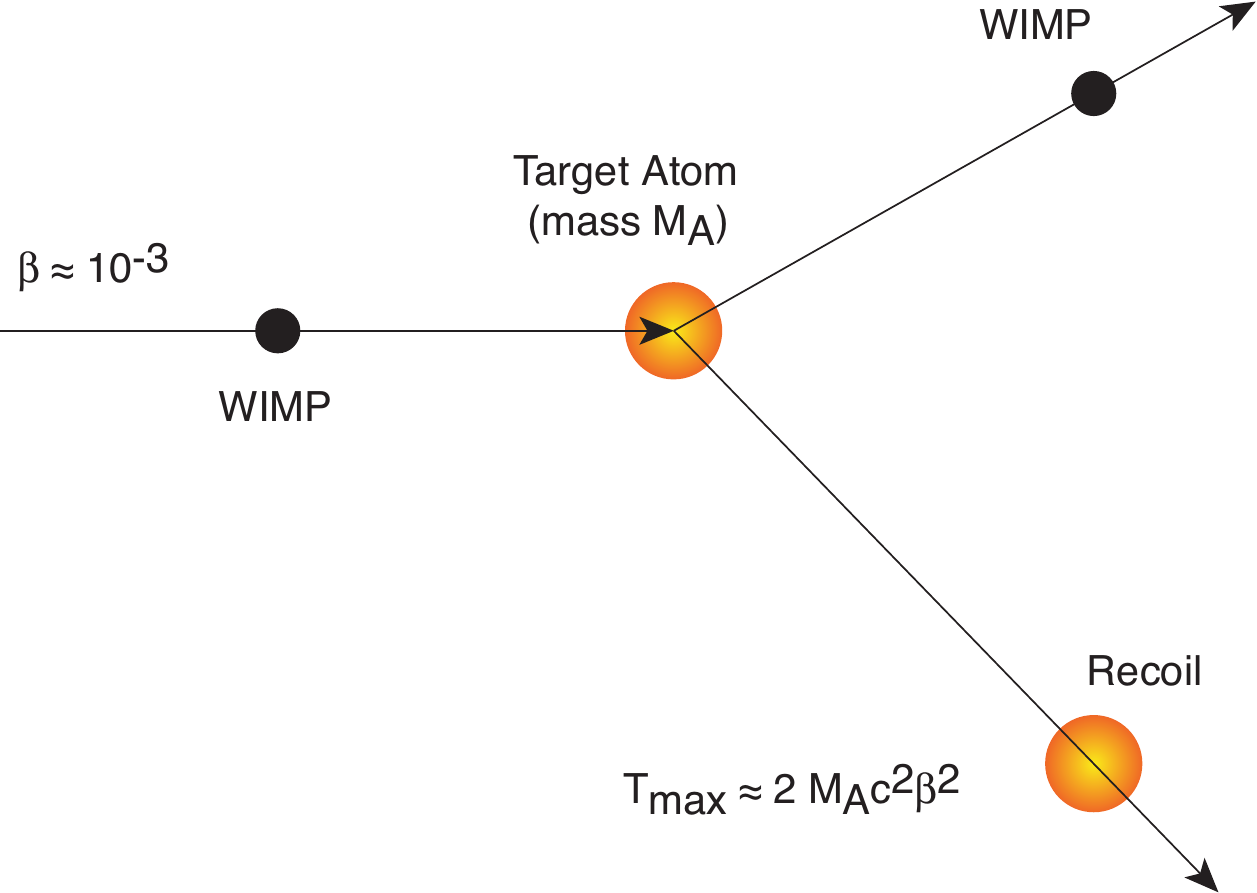}
\includegraphics[width=70mm,height=40mm]{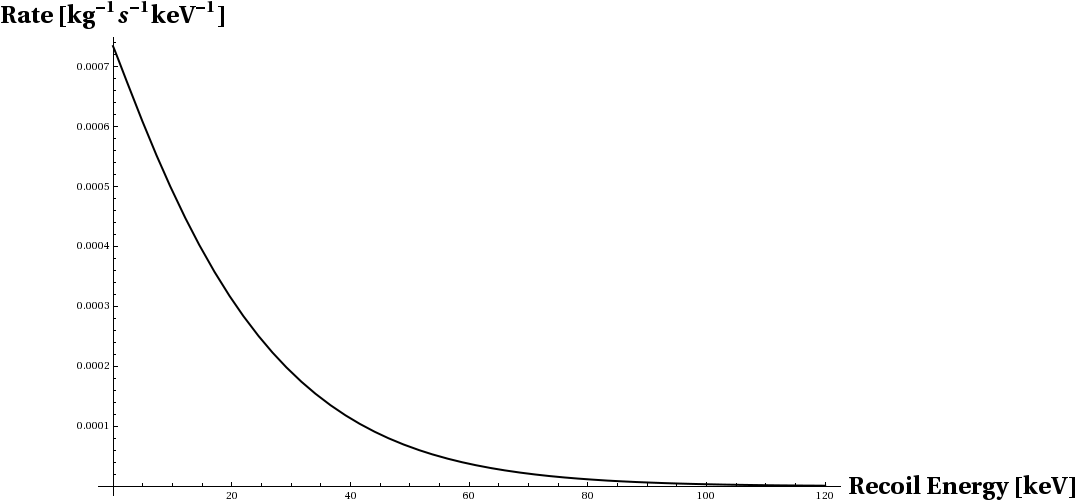}
 \caption{Left, a diagram showing the idea of a WIMP interaction in a detector. Right, the expected event rate per kg per second per keV in argon for a WIMP of the mass of 60 GeV/c$^2$ and standard halo parameters. }
 \label{fig:recoil}
\end{figure}

The direct detection experiments focus on detecting the WIMPs which should be present in the solar neighborhood as a result of the presence of the galactic halo surrounding the Milky Way. It is generally assumed that the WIMPs in the halo behave according to the Standard Halo Model, which states that the Dark Matter particles behave like a gas with a Maxwell-Boltzmann velocity distribution \cite{SHM}. If this is the case, then the bulk velocity of the WIMPs with respect to the galaxy rest frame would be equal to zero. However, due to the gravitational pull of the center of the galaxy, the galactic disk and the solar system with it rotate around the galactic center. The velocity of this movement for the Sun is about 220 km/s. Another effect is that of the annual voyage of the Earth around the Sun, with a speed of 48 km/s which should be added to the solar system speed. Together, these speeds become the effective speed of the WIMP wind with respect to a detector on Earth. WIMPs interacting in a detector (Fig. \ref{fig:recoil} left) will have a kinetic energy which will be the result of this speed. The resulting recoil energy transfers should be of the order of 10-100 keV, where more events are expected at low energies as can be seen in Fig. \ref{fig:recoil} right, where a recoil spectrum for a WIMP of 60GeV/c$^2$ mass in argon is shown. From this dependence it can be seen that the lower the energy threshold the more events a detector can see.  Apart from this, one should remember that the expected WIMP event rates are very small, with the current experimental limits it is expected to see less than 0.01 evt/kg/day in a detector. So, in order to operate a Direct Dark Matter search experiment the requirements are a large mass, a low energy threshold (implying an excellent control of the detector background) and, finally, patience.

The background rejection is usually divided into discriminating the more abundant background coming from $\beta$ and $\gamma$ interactions and that from neutrons which in principle are much more dangerous since they can mimic a WIMP signal. 

\subsection{Electron and Gamma Backgrounds }
\begin{figure}[htp]
 \centering

\includegraphics[width=70mm]{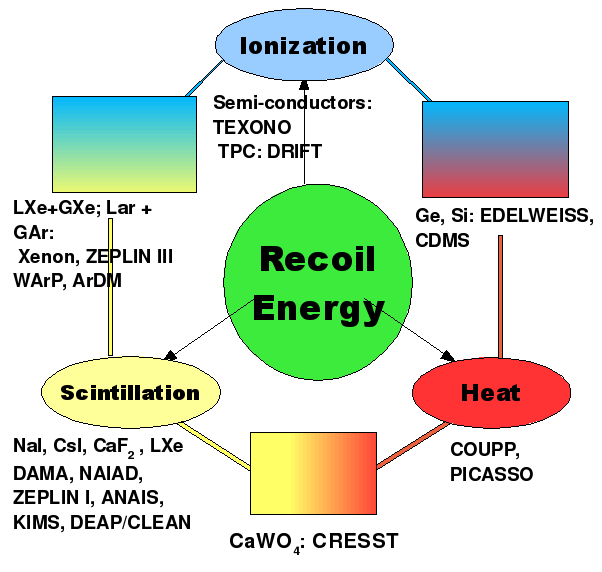} 

 \caption{The three main interaction channels of Dark Matter and a selection of the experiments that use them in their search for Dark Matter. Experiments found halfway between the interaction channels use both of these channels.}
 \label{fig:daq_setups}
\end{figure}

It is expected that a WIMP interacting with a detector atom should interact principally with the nucleus transferring its kinetic energy and causing the detector atom to recoil. This is caused by the fact, that it has no electric charge (otherwise it would have been already detected) which would allow it to interact with the electron shell. This leads to expected differences of these interactions from those of minimum ionizing particles, which allows the suppression the $\beta$ and $\gamma$ (electron-like) background using clever detection techniques. Most suppression techniques in some way are based on the difference in ionization density between electron-like background events and recoil-like events caused by the interaction of a WIMP or neutron with a detector nucleus. If one is able to measure the total energy and the ionization (or scintillation, which is complementary) of a given interaction it is possible to deduce if the interaction was caused by $\gamma / \beta$ background type, or one of the recoil-like interesting events. In order to use these background rejection techniques usually, but not always, the interaction needs to be registered in more than one out of the scintillation, ionization and heat (a result of the of phonons in a crystal lattice) channels. The break down of some experiments based on their chosen data acquisition methods can be seen in Fig. \ref{fig:daq_setups}, where the experiments found between two interaction channels acquire both of them. Currently the most rapidly developing groups of detectors are the cryogenic crystal detectors i.e. (CDMS, CRESST, Edelweiss) and liquid noble gas detectors (i.e. XENON, WArP, CLEAN, ArDM).

\subsection{Neutron Background}
Neutron interactions are a much more dangerous background for Dark Matter experiments since, as said before, they can mimic the WIMP interactions. This means that the above mentioned discrimination methods cannot be used to reject neutron events. This is because, the neutrons, like WIMPs, have no electric charge and so should interact only with the nuclei of detector atoms. There is a difference however and this is of the interaction cross-section - the neutrons are much more likely to interact than WIMPs. And if one expects a WIMP to interact rarely in the detector, a neutron will interact almost always, and what is more, for a large enough detector it should interact more than once. This gives another strong reason to scale up the detector masses. Another possibility is of course trying to determine the expected neutron rate in the detector by shielding and measuring the detector components and surroundings for the expected neutron flux (usually coming from uranium and thorium daughters' interactions) and by having under control the muon flux, which can also cause neutron interactions in the detector. A large effort is being pursued to construct future detectors from ultra low background materials to avoid spurious neutron events as much as possible.

\subsection{Dark Matter Experimental Signatures}

Apart from rejecting all known interactions and looking for what remains, there is another way to approach the direct detection challenge and that is to search for specific experimental signatures which we expect that Dark Matter should present. There are two such signatures that are well known and searched for experimentally, namely the annual and diurnal modulations. The annual modulation is a result of the yearly movement of the Earth around the Sun. The speed of the Earth can either augment or decrease (June or December, respectively) the effective speed of the detector versus the halo rest frame, hence changing the expected number of events in the detector. In effect, one should observe a sinusoidal behaviour with a period of 1 year and a peak in June. As of now the experiments searching for this effect are usually crystal scintillator detectors with a large mass like DAMA-LIBRA, which will be discussed later.

\begin{figure}[htp]
 \centering
\includegraphics[width=70mm]{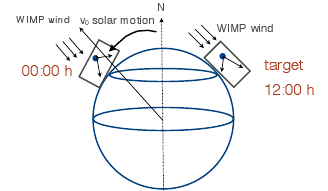}  
 \caption{The diurnal modulation detection principle \cite{diurnal}.}
 \label{fig:modulations}
\end{figure}

The diurnal effect results from the rotation of the Earth around its axis. Due to this movement and the tilt of the Earth with respect to the galactic plane the direction of the effective speed of the WIMP wind changes by 90 degrees \cite{directionalDM}, as in Fig. \ref{fig:modulations}. This is an effect that should be extremely hard to imitate by background events, however to detect it a detector capable of measuring particle tracks is needed and this is a very difficult experimental challenge given the expected low energy of WIMP interactions. Nevertheless, such an experimental effort is on the way and low pressure gas TPCs are being tested and operated by various collaborations.    

\subsection{Cryogenic Crystal Detectors}

\subsubsection{CDMS II Result}

\begin{figure}[htp]
 \centering
 \includegraphics[width=50mm]{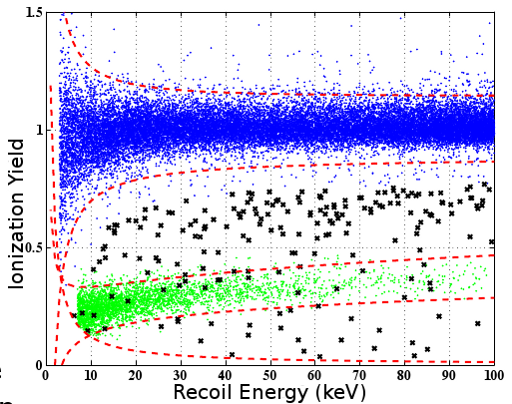}  
\includegraphics[width=50mm]{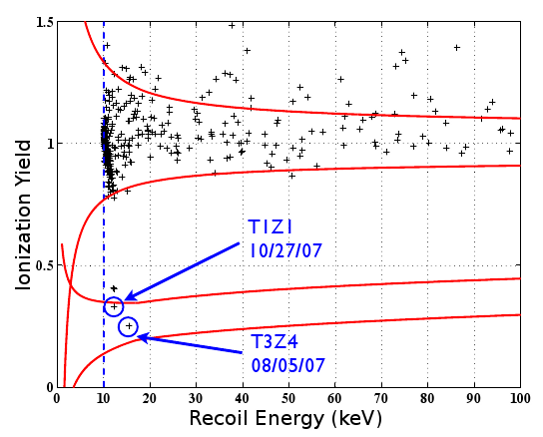}
 \caption{Left, the calibration data from the CDMS II experiment - the upper band are the bulk electron recoils, the lower band are neutron recoils and the black squares are electron surface events removed with the phonon timing cut. Right, the result of the CDMS II data run, two events are observed in the nuclear recoil band \cite{CDMS_?_res}.}
 \label{fig:CDMS}
\end{figure}

The CDMS II (Cryogenic Dark Matter Search) experiment was located in the SOUDAN underground laboratory and has been operating there since June 2006.
It consisted of 30 crystal detectors totalling a mass of 4.75 kg of Ge and 1.1 kg of Si kept at temperatures lower than $50$mK. This allowed to read out both charge and phonons, as heat,
which results in very good background suppression. A timing parameter is added to the discrimination based on the charge/heat ratio to exclude events close to the surface of the detectors, where the standard discrimination method can malfunction. The details can be seen in Fig. \ref{fig:CDMS}.

The recent result of CDMS II, which announced their analysis of the total data sample on December 17th 2009 has surely been the most anticipated and discussed in the last month of the year. Just before the CDMS II talks a lot of excitement was spreading in the community, because of gossip that CDMS II may have seen Dark Matter. At the end of the day, this was not confirmed, however, for the first time the CDMS collaboration has seen Dark Matter candidate events. 

The analysis was based on an exposure of 612 raw kg-days corresponding to 194.1 kg-d WIMP equiv. @ 60 GeV/c$^2$ WIMP mass (due to fiducial cuts) in the 10 -100 keV recoil energy range. Two events were observed, while the expected background was $0.8 \pm 0.1 (stat) \pm 0.2 (syst)$ \cite{CDMS}. Both of the events could raise some doubts and since there is a 23\% probability that both events could be caused by background a Dark Matter detection claim could not have been made. However it may be considered as another sign, in the opinion of the author, that Dark Matter may be just around the corner. 

\subsubsection{Other Cryogenic Detectors}

CDMS is not the only cryogenic detector using the heat readout technique. Other noteworthy efforts are EDELWEISS \cite{EDELWEISS} based in the underground laboratory of Modane and CRESST based at Gran Sasso Laboratory \cite{CRESST}. EDELWEISS is an experiment analogous to CDMS, in that it uses germanium and charge is read out in parallel with heat. CRESST on the other hand uses CaWO$_4$ crystals as the detector medium and register scintillation together with phonon vibrations of the crystal lattice. Both experiments have recently published experimental Dark Matter search results, however their exposure is as of yet smaller than that of CDMS (EDELWEISS: 144  kg days \cite{EDELWEISS}, CRESST: 42 kg days. \cite{CRESST}).

\subsection{Scintillation Detectors}

\subsubsection{DAMA and DAMA-LIBRA}

\begin{figure}[htp]
 \centering
 
\includegraphics[width=66mm]{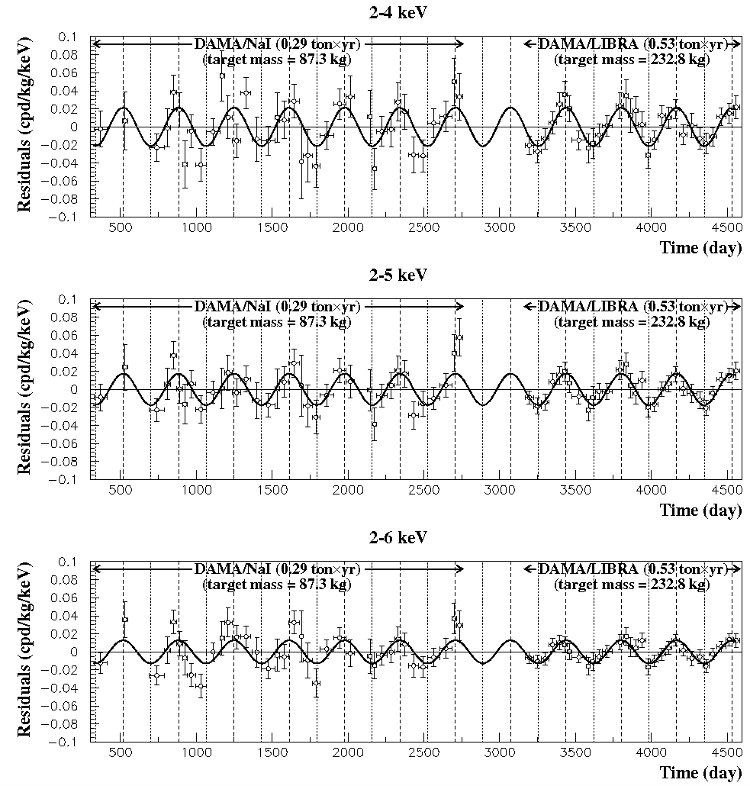}

 \caption{The annual modulation effect observed by the DAMA detector and its successor DAMA-LIBRA \cite{DAMA_LIBRA}.}
 \label{fig:DAMA}
\end{figure}

The DAMA experiment and its successor DAMA-LIBRA is currently the experiment with the largest experimental exposure totaling 0.82 ton years. It is placed in the underground laboratory in Gran Sasso, Italy. The detection technique is based on registering scintillation pulses from NaI(Tl) crystals (100 kg for the DAMA phase and 250 kg for the DAMA-LIBRA phase) specially constructed to achieve low internal background and shielded from outside radioactive sources. The idea behind DAMA is to search for the annual modulation effect looking at the total number of interactions registered in the detector throughout the year. In fact, DAMA has made the claim of discovery of Dark Matter based on their observation of the annual modulation first with the DAMA detector \cite{DAMA} and then confirmation with the DAMA-LIBRA apparatus\cite{DAMA_LIBRA}. The total modulation is compatible with a period of one year peaking in June, as can be seen in Fig. \ref{fig:DAMA}. The DAMA result has been known since the year 2000, however since then no experiment has been able to confirm the DAMA result \footnote{Two months after this talk, a preprint published by the CoGeNT collaboration suggests that they have observed an excess of events in th energy region close to that of DAMA \cite{CoGent}. This result is being tested by the collaboration.}. What is more, most experiments seem to exclude the DAMA claim for the most standard WIMP and halo models. 

\subsubsection{Testing DAMA}
It has to be said that the DAMA result has not been as of yet thouroughly tested by an analogous crystal scintillator experiment. Some attempts were made, for instance NAIAD \cite{NAIAD} ran from 2000 to 2003 and used pulse shape discrimination in a NaI detector. It acquired 44.9 kg years of exposure, but was unable to exclude the „DAMA region” \cite{NAIAD}. The KIMS detector - located in Korea (Yangyang) is taking data with 12 CsI detectors (total 104.4 kg) and has so far acquired 3409 kg-days of exposure. Their observations are up to now consistent with a null result \cite{KIMS}. Another group, the ANAIS collaboration, will try to repeat the DAMA measurement with specially prepared NaI crystals. At the time of the talk, the first 9.7 kg module was being tested at
the Canfranc underground laboratory. The collaboration is planning to operate a total of $\sim$150 kg of NaI and look for the annual modulation signal \cite{ANAIS}. 
Another experimental effort has been recently announced by the Princeton group \cite{GalbiatiatWIN}, which has begun an R\&D program on development of NaI crystals radioclean from $^{40}$K, a radioactive contamination present in sodium which emits radiation exactly around 2-4 keV where DAMA has seen their signal.
Inclusion of a 4$\pi$ liquid scintillator veto would reduce $^{40}$K by many orders of magnitude therefore eliminating one of the last serious doubts concerning the DAMA result.

\subsection{COUPP - the Renaissance of the Bubble Chamber}

The COUPP experiment is a rather new idea that has been proposed only a few years ago, yet it has already been able to provide significant physics results. It is based on the concept of the bubble chamber which has been used in many successful experiments in the last century. What makes it very interesting from the point of Dark Matter detection is that the bubble nucleation depends on the dE/dx energy deposition which is low for minimum ionizing particles and high for recoil-like events. Running at low pressure it is possible to fine tune the chamber in such a way that only recoil-like events create bubbles in the chamber, hence suppressing the electron/gamma background. It is also possible to use different liquids as the target medium, which gives the possibility to cross-check possible systematics and explore different effects like the dependence of the interactions on the spin of the detector atoms.

The event readout is performed by means of cameras which observe the target volume and register emerging bubbles. Neutrons, as in most experiments, can be excluded by observing multiple interactions in the detector. COUPP has published the results of a test run with a 1.5 kg chamber \cite{COUPP_result}, which already was one of the most competitive spin-dependent exclusion plots.

\subsection{Liquid Noble Gas Detectors}

In recent years liquid noble gasses  detectors have become more and more popular as a Dark Matter search medium. The reasons are mainly their relative ease of operation and their relatively low price which together allow for an easier and cheaper construction of heavier detectors and last but not least their intrinsic scintillation properties which allow very good background discrimination. 

\begin{figure}[htp]
 \centering
 \includegraphics[width=70mm,height=64mm]{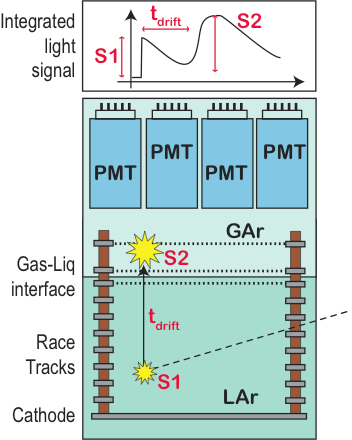}
 \caption{The detection principle in a two-phase liquid noble gas detector, based on the WArP experiment \cite{WArP}.}
 \label{fig:double_phase}
\end{figure}

\begin{figure}[htp]
 \centering
  \includegraphics[width=90mm]{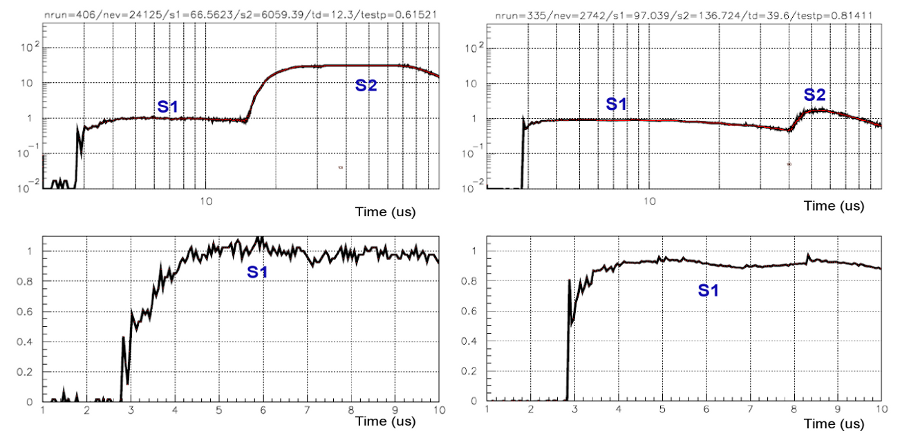}
 \caption{Background discrimination in liquid noble gases using the example of liquid argon. Upper plots show the S2 and S1 pulses for an electron-like (upper left) and recoil-like(upper right) events. Bottom plots show a close up of the primary pulses for the respective events. Note the fast rise of the recoil-like event (bottom right) \cite{WArP_some}. }
 \label{fig:dphase_discrim}
\end{figure}

Liquid noble gas detectors are operated in either double or single phase. An example of a double phase detector can be seen in Fig. \ref{fig:double_phase}. The detection method is based on registering the primary scintillation pulses (S1) coming
  from the light emitted just after the interaction in the liquid phase and the secondary pulses (S2), i.e. scintillation light emitted by ionization
  electrons, which are first drifted up in an electric field and then accelerated in the
  gaseous phase. This technique provides a very strong rejection power due to the fact that the scintillation to ionization ratio in noble liquids
again depends on the dE/dx of the incident particle \cite{S2/S1}. Therefore the
   discrimination between $\gamma$/$\beta$ particles and those interacting
   with the nucleus i.e. WIMP and neutrons can be conducted by
   comparing their S2/S1 ratio (Xe, Ar, Ne), see Fig.\ref{fig:dphase_discrim}. Another method which is possible to use in liquid noble gases is the S1 pulse shape discrimination technique which utilizes the fact that their scintillation light is emitted via the dimerization process. Usually there are two different scintillating molecular states, of which the relative density depends again on the dE/dx of the incident particles \cite{pulse_shape}. This causes a difference in the shape of the primary scintillation signal as in Fig.\ref{fig:dphase_discrim}. This effect can be practically used only in argon and neon, because in xenon the decay times of the molecules are too close together. 

The combination of the above mentioned methods allows for the discrimination of the more abundant $\gamma$/$\beta$ background, which is especially important in argon, because it is dominated by the background coming from the cosmogenic isotope of $^{39}$Ar \cite{WARP_Ar}. To
  operate in a Dark Matter
  search the background
  suppression must be good
  enough to exclude these events. Another solution is to use
  isotopically depleted argon (centrifuges or
  underground reservoirs \cite{WARP_Cris}). 
 Neutrons require a veto detector or a large enough proper detector.

Single phase detectors do not use the charge readout in the gaseous phase and therefore must depend on self-shielding which is best in Xe detectors ($\sim~$3 g/cm$^3$ density) or pulse shape discrimination. Most projected single-phase detectors use a 4$\pi$ detector coverage which allows for event localization and therefore detector fiducialization.

\subsubsection{XENON}
The XENON Experiment is being operated
  at the Gran Sasso laboratory. The collaboration uses the two-phase technique with xenon as the target medium. The first
  detector, XENON 10, published DM
  search results in 2008 \cite{XENON} and this result continues
  to be one of the leading experimental limits. 10 events were observed, with 6.8
  expected from $\gamma$ leakage. At the time of the presentation, the next phase, XENON100 was in commissioning at
  LNGS. Currently it is taking data, and  with the 50 kg fiducial volume it should have reached CDMS II exposure after only weeks of data taking.

%

\subsubsection{WArP}

The WArP collaboration is operating a 100l detector in the Gran Sasso laboratory, together with a small 2.3 liter prototype detector. The collaboration uses the double phase technique and is the first experiment to publish results of a Dark Matter search with liquid argon \cite{WARP_DM}. The main, 100l detector has been
  commissioned at LNGS in
  May 2009. Due to problems with the high voltage cable,  the detector was unable to run in double phase, and so it was opened in
  August. During the operation of the chamber
  in single phase the DAQ and
  reconstruction algorithms were
  tested. Subsequently, the new HV system
  was tested and the detector
  should begin operation in the
  spring of 2010 \cite{Claudio}.

\subsubsection{DEAP/CLEAN}

The DEAP/CLEAN collaboration is
   planning to operate single phase
   argon and neon detectors, using
   only the pulse shape discrimination
   capabilities of these detectors in the underground laboratory in SNOlab \cite{CLEAN_SNO}.

The MiniCLEAN detector with 150 kg of
   Fiducial mass is planned to be
   commissioned at SNOLAB in the middle of
   2010. In 2011 the DEAP-3600 detector with a
   fiducial mass of 1t is planned to be
   commissioned. At the time of the talk it was still in design
   phase \cite{DEAP}.

\subsubsection{ArDM}

ArDm is a 1 ton double phase argon
   detector. However, the charge will be collected 
   directly using LEM (Large Electron Multiplier) detectors instead of using PMTs to register secondary scintillation.
The ArDM chamber was already operated at
   ground level with half of the
   PMTs. A light yield of
   0.5 phe/keV was obtained, consistent with the expectations of 1phe/keV with all PMTs.
The possible locations of the detector are SUNLAB    (Poland), Canfranc (Spain) and
   Slanic (Romania). The detector is described more thoroughly in \cite{Haranczyk}.

\subsection{Directional Detectors}
  The searches for diurnal modulation effects are being conducted using low pressure gas TPCs. Some projects include DRIFT \cite{DRIFT}, DMTPC \cite{DMTPC}. A review of such experiments can be found in \cite{direc_review}.

\subsection{Axion Searches }
There is also an ongoing effort to detect axions as a source of Dark Matter. Examples are the CAST \cite{CAST} and ADMX \cite{ADMX} experiments.

\subsection{Current Experimental Status}

\begin{figure}[htp]
 \centering
 
  \includegraphics[width=81mm]{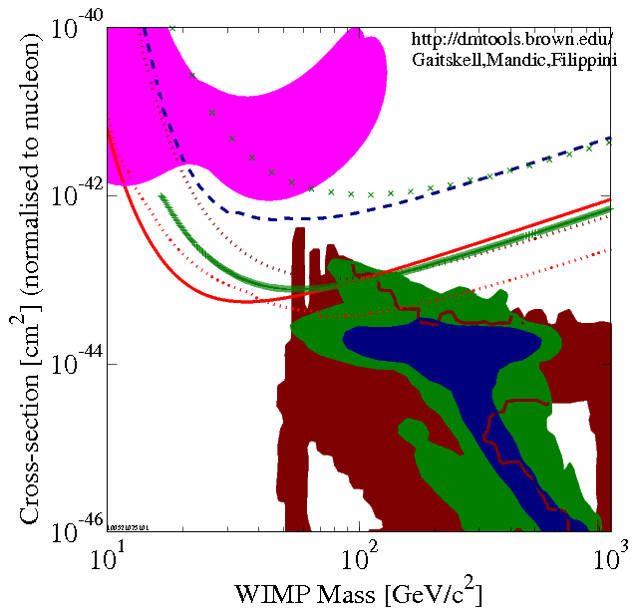}
\includegraphics[width=51mm]{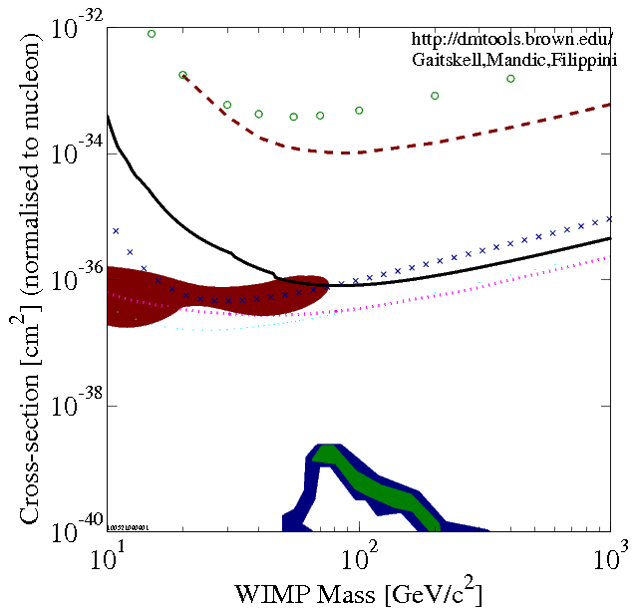}
\includegraphics[width=51mm]{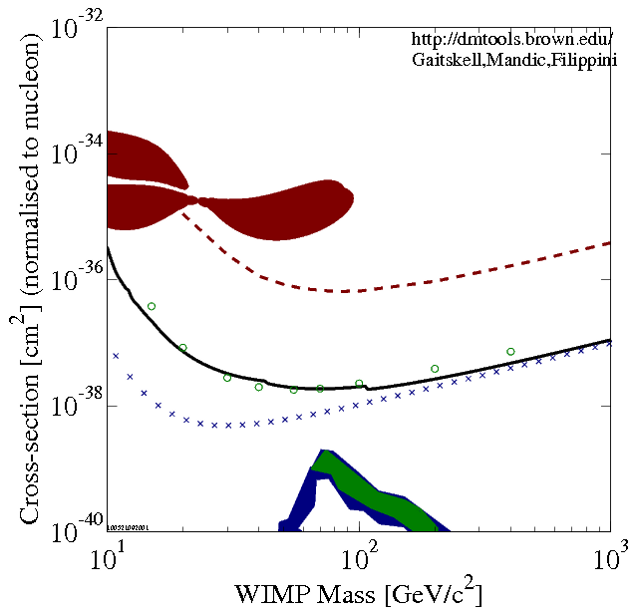}

 \caption{The exclusion plots in spin-independent case (top), in descending order: for WArP (crosses), CRESST (dashed), EDELWEISS (dashed dots), ZEPLIN (crosses), XENON 10 (line), and CDMS II (dots) below are the predictions of SuperSymmetry, the area above is the region preferred by the DAMA result. Bottom plots show the spin-dependent cases for coupling with protons (left) for ZEPLIN III (circles), EDELWEISS (dashed), CDMS II (line), XENON 10 (crosses), COUPP (dots), PICASSO (points) and neutrons (right) EDELWEISS (dashed), ZEPLIN III (circles), CDMS II (line), XENON 10 (crosses) both show also the DAMA preferred region.  \cite{Dendera}}
 \label{fig:exclusion}
\end{figure}

The current experimental situation is shown in Fig. \ref{fig:exclusion} where the exclusion plots for the leading experiments are presented. These plots show the WIMP mass - interaction cross-section parameter range disallowed by null results of the direct Dark Matter searches. The predictions of SuperSymmetric models are also plotted. It can be seen that especially in the spin-independent case Fig. \ref{fig:exclusion} top, the experimental results of CDMS II, XENON 10 and ZEPLIN III are already on the edge of the predicted region and what is more the DAMA result is almost completely ruled in out in all cases for the standard WIMP and halo scenario. In the spin-dependent case, Fig. \ref{fig:exclusion} bottom left and bootom right for proton and neutron coupling respectively the results are still quite far from the theoretical predictions. However, also in this case the DAMA preferred region is on the verge of being excluded.

\subsection{What the (Nearest) Future Holds}
According to the predictions of the experimental collaborations 2010 should be a very interesting year in Dark Matter detection. Tab. \ref{tab:sens} presents the lowest sensitivities that are expected to be reached in 2010. One should underline that these are the maximum sensitivity values possible to obtain only for the optimal WIMP mass for a given detector - for other masses the sensitivity would not be as good. 

If one is to believe these declarations it would mean that in the course of the year 2010 more than one detector should reach a maximum sensitivity of 10$^{-45}$ cm$^2$ Spin Independent (SI) which would put the experimental results below the area most probable from the point of view of theoretical expectations. So it would seem that apart from experimental hints there is also a suggestion from theory that the WIMP should be discovered soon.

\begin{table}[htp]
 \centering

\begin{tabular}{|l|l|l|l|}
\hline Detector & Mass & max. sensitivity & months run  \\  \hline
WArP & 144 kg LAr & $\sim$ 5$\cdot$ 10$^{-45}$ cm$^2$  & 3  \cite{Segreto}  \\ \hline
XENON 100 & 50 kg Xe (fiducial) & $\sim$ 2$\cdot$ 10$^{-45}$ cm$^2$ & 7  \cite{Ferrella}  \\  \hline
CDMS II & 15 kg Ge & $\sim$ 4$\cdot$ 10$^{-45}$ cm$^2$ & - \cite{CDMS_?_res}  \\ \hline
CLEAN & 150 kg LAr (fiducial) & $\sim$ 7$\cdot$ 10$^{-46}$ cm$^2$ & 24  \cite{DEAP}  \\ \hline
LUX & 100 kg LXe (fiducial)  & $\sim$ 7$\cdot$ 10$^{-46}$ cm$^2$  & 10  \cite{McKinsey_TAUP}  \\ \hline
XMASS & 100 kg LXe (fiducial) & $\sim$ 3$\cdot$ 10$^{-45}$ cm$^2$ & - \cite{XMASS_talk}  \\ \hline
COUPP & 60 kg Bubble Chamber & $\sim$ 5$\cdot$ 10$^{-39}$ cm$^2$ SD & - \cite{COUPP_TAUP}  \\ \hline
 \end{tabular}

 \caption{The expected maximum sensitivities of the detectors expected to run in 2010 along with the time needed to obtain it (if given). All predictions are for Spin Independent measurements except COUPP. }
 \label{tab:sens}
\end{table}

\section{Conclusion}
The field of Dark Matter searches is a very difficult but exciting field of astrophysics. There is an enormous experimental effort going on and one can hope that it will bear fruit in the nearest future, which could change our outlook both on astrophysics and particle physics.

\end{document}